\documentclass[fleqn,11pt]{wlscirep}
\usepackage{amsmath}
\usepackage{amsfonts}
\usepackage{amssymb}
\usepackage{color}
\usepackage{graphicx}
\usepackage{array}
\usepackage[export]{adjustbox}
\usepackage{float}

\title{Griffiths phases in infinite-dimensional, non-hierarchical modular networks}

\newcommand{\av}[1]{\langle{#1}\rangle}
\newcommand{\kupp}{k^\mathrm{_{[upp]}}}
\newcommand{\klow}{k^\mathrm{_{[low]}}}
\newcommand{\kout}{k^\mathrm{_{[out]}}}

\newcommand{\Smin}{S_\mathrm{min}}
\newcommand{\expS}{\phi}

\author[1,*]{Wesley Cota}
\author[2]{G\'{e}za \'{O}dor}
\author[1,3]{Silvio C. Ferreira}
\affil[1]{Departamento de F\'{\i}sica, Universidade Federal de Vi\c{c}osa,
	36570-000, Vi\c{c}osa, Minas Gerais, Brazil}
\affil[2]{MTA-EK-MFA, Centre for Energy Research of the Hungarian
Academy of Sciences, H-1121 P.O. Box 49, Budapest, Hungary}
\affil[3]{National Institute of Science and Technology for Complex Systems, Rio de Janeiro, Brazil}

\affil[*]{Correspondence and requests for materials should be addressed to \href{mailto:wesley.cota@ufv.br}{wesley.cota@ufv.br}.}

\begin{abstract}
Griffiths phases (GPs), generated by the heterogeneities on modular networks,
have recently been suggested to provide a mechanism, rid of fine parameter
tuning, to explain the critical behavior of complex systems. One conjectured
requirement for systems with modular structures was that the network of modules
must be hierarchically organized and possess finite dimension. We investigate
the dynamical behavior of an activity spreading model, evolving on heterogeneous
random networks with highly modular structure and organized non-hierarchically. We
observe that loosely coupled modules act as effective rare-regions, slowing down
the extinction of activation. As a consequence, we find extended control
parameter regions with continuously changing dynamical exponents for single
network realizations, preserved after finite size analyses, as in a
real GP. The avalanche size distributions of spreading events exhibit robust
power-law tails. Our findings relax the requirement of hierarchical organization
of the modular structure, which can help to rationalize the criticality of
modular systems in the framework of GPs.
\end{abstract}
\begin{document}

\maketitle

\section{Introduction}

A recurrent feature of complex systems is the presence of critical states, in
which spatial and temporal correlations diverge\cite{Sornet2006,Tauber2014}. 
A fundamental question is why and how a complex system would be tuned to
criticality\cite{SOC,Dickman2000,Chialvo2010}. As a very important example,
recent experimental evidences suggest that the brain operates near
criticality\cite{Beggs2003,Chialvo2010,Haimovici2013,plenz2014criticality}. 
Information processing capabilities, sensitivity and the dynamic
range of stimuli, where the collective response varies significantly,
are optimal in this region\cite{LM07,sporns2010networks,Larremore2013,Beggs2012}.
Simple models on homogeneous substrates\cite{Chialvo2010,Kinouchi2006} have 
frequently been used to answer this question and criticality is often
associated with some self-organization\cite{SOC} or evolutionary selection
mechanism\cite{adap}. However, heterogeneity of the networks mediating the
interactions among the agents of a dynamical process can be relevant for the
outcomes of models investigated on
them\cite{PastorSatorras2015,Castellano2009,Rodrigues2016}, in particular, the
quasi-static (quenched) disorder, with timescales much longer than those of the
dynamics. Thus, it is a challenge to understand how quenched disorder originated
from the heterogeneous network topology affects the observed critical state.

\begin{figure}[t!]
	\centering
	\includegraphics[width=0.6\linewidth]{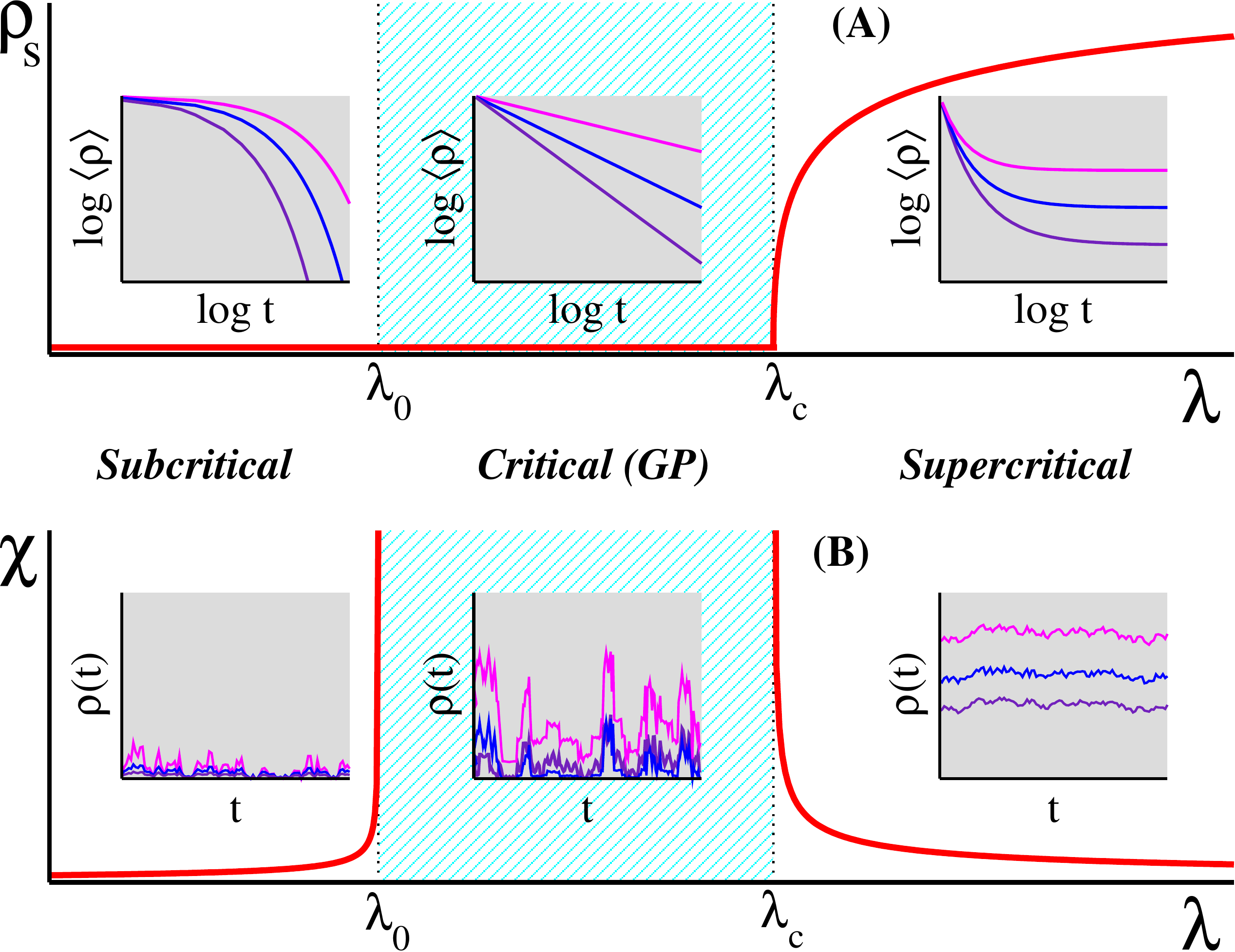}
	\caption{{Griffiths phases and dynamical criticality in spreading
			processes}. The top panel shows the stationary order parameter $\rho_\text{s}$,
		which is the average density of active vertices, against the control parameter
		$\lambda$.  The order parameter has a finite value above the critical point
		$\lambda_\text{c}$ and vanishes as $\lambda\rightarrow\lambda_\text{c}$.  The
		stationary density is zero for $\lambda<\lambda_0$. In both regimes, the
		characteristic times to reach the stationary state are finite, typically given
		by exponential decays in the curves of average density against time.  The asymptotic
		density is still null in the critical region, but the approaching to the
		asymptotic value is slow, typically a power-law time decay, manifested by
		straight lines on log-log plots. The bottom panel shows the stationary dynamical
		susceptibility\cite{Mata2015,Cota2016}, defined as the relative variance of the
		order parameter, given by $\chi=N[{\av{\rho^2}-\av{\rho}^2}]/{\av{\rho}}$,
		against the infection rate under the presence of a small external stimulus,
		which can be a spontaneous self-activation\cite{Sander2016}. Sub and
		supercritical phases present finite susceptibility diverging as we
		approach  the critical region, in which it remains infinite. The fluctuations
		are finite in the  off-critical interval and huge within the  whole critical
		region, representing high sensitivity to external stimuli. The subcritical,
		critical and supercritical regimes are schematically represented in the left,
		center, and right insets, respectively.}
	\label{fig:GPcheme}
\end{figure}

In condensed matter physics, quenched disorder can lead to the so-called
Griffiths phases (GPs)\cite{Griffiths1969} with dynamical criticality and high
sensitivity to external stimuli in an extended parameter space\cite{Vojta2006b}.
Dynamical criticality means long-term temporal correlations that imply slow
relaxation and broad distributions of interevent times manifested usually as
power laws (PLs)\cite{Tauber2014,Henkel2008}. GP is the consequence of rare
regions (RRs), consisting of local (sub-extensive) supercritical (active)
domains, which occur with small probability but sustaining activity for long
times (exponential in domain size). To understand GPs in critical dynamics,
consider a dynamical spreading process with active and inactive states, a
control parameter $\lambda$ and an order parameter $\rho$ (density of active
individuals) determining the system phase. Inactive states are also called
``absorbing'' because, once visited, no other state can be reached from it
without an external source~\cite{Henkel2008}. The system is in a globally active
phase for $\lambda>\lambda_\text{c}$, with a non-zero order parameter as
$t\rightarrow\infty$, or in an inactive one, for $\lambda<\lambda_0$ with
$\rho=0$, for $t\rightarrow\infty$. Long lived RRs are absent in the latter
case\cite{Vojta2006b}. Response functions, which quantify the sensitivity to
external stimuli, are finite in both ranges; see Fig.~\ref{fig:GPcheme}. In the
interval $\lambda_0 < \lambda<\lambda_\text{c}$, the activity in RRs is long
lived, but ends up due to the fluctuations in the finite sized local patches.
Convolution of low-probability RRs and exponentially long lifetimes results in a
slow relaxation and highly fluctuating dynamics characterized by nonuniversal
exponents in this interval, constituting a GP~\cite{Vojta2006b}. Response
functions become very large (formally infinite in the thermodynamic limit) in
this range and the system exhibits hypersensitivity to external stimuli. 
Figure~\ref{fig:GPcheme} shows a scheme for GPs and dynamical criticality.  A central
question is if the RR effects are strong enough to alter the phase
transition\cite{Barghathi2014}.

Critical systems can be sensitive to quenched disorder when their dimension is
sufficiently low\cite{Vojta2006b}. A hypothesis based on activity spreading
models claims that the heterogeneity effects become irrelevant in the
thermodynamic limit in case of infinite-dimensional random
graphs\cite{Munoz2010}. The dimension $d_g$ of a graph is given by the relation
between the average number of nodes $\mathcal{N}$ enclosed in a distance $l$, such that
$\mathcal{N}\sim l^{d_g}$. For small-word networks, for which distances increase
logarithmically with the graph size\cite{Albert2002}, we formally have
$d_g=\infty$. {Some evidences suggest that the brain network} is {heterogeneous
	and } present hierarchical modular organization, in which modules are themselves
composed by modular substructures at distinct
levels\cite{sporns2010networks,Bullmore2009,Meunier2010}. These inspired Moretti
and Mu\~noz\cite{Moretti2013} to investigate activity spreading models on
hierarchical modular networks of finite dimension, for which GPs and extended
critical regions were observed (see also \cite{Odor2015,Li2017})  and to
conjecture that the brain criticality could be effected by quenched disorder
without fine parameter tuning. It does not mean that one cannot find relevant
effects in finite non-modular systems\cite{Cota2016,Lee2013,Mata2015}. 
Moreover, long-range connections can drastically increase the network dimension,
even if they constitute just a small portion of the
graph\cite{Newman1999,Barrat2000}. The empirical organization of biological
networks  is highly complex and subjective\cite{sporns2010networks} and,
therefore, it is not {completely} clear whether {real brain networks} in a
cellular level are actually hierarchical\cite{Hilgetag2016}. Furthermore,
modular graphs without hierarchical structure are observed in diverse important
systems such as socio-technological\cite{Ebel2002,Palla2007} or protein
interaction networks \cite{protein-data}, but the existence of extended critical
regions due to the quenched disorder on such systems has not been considered
extensively.

To our knowledge, no investigation has been done to scrutinize whether hierarchy
is really a necessary condition for the emergence of GPs. The present work aims
at to fill this gap, using simulations of activity spreading models on
non-hierarchical modular structures. Recently, optimal fluctuation
theory\cite{Lee2013} and simulations provided extended critical regions on
heterogeneous networks of finite size constrained to averages over independent
network samples\cite{Cota2016}. This inspired us to investigate the dynamical
behavior of the continuous time Markovian susceptible-infected-susceptible (SIS)
model, which has been used to describe activity or information spreading in
socio-technological and biological
systems\cite{Moretti2013,Odor2016Conn,Kitsak2010,Castellano2017}, on loosely
coupled network of modules.  We found extended control parameter regions with
non-universal PL decays of activity in time, which are size-independent, calling
for the existence of real GPs in infinite dimensional, but loosely connected
modular structures. Thus, our results point out that we can relax the
requirement of hierarchical organization and
large-world\cite{Munoz2010,Moretti2013} for the existence of GPs on modular
networks, although these factors certainly enhance RR effects.

\section{Results}

\subsection{Synthetic modular networks}

\begin{figure}[b!]
	\centering
	\includegraphics[width=0.72\linewidth]{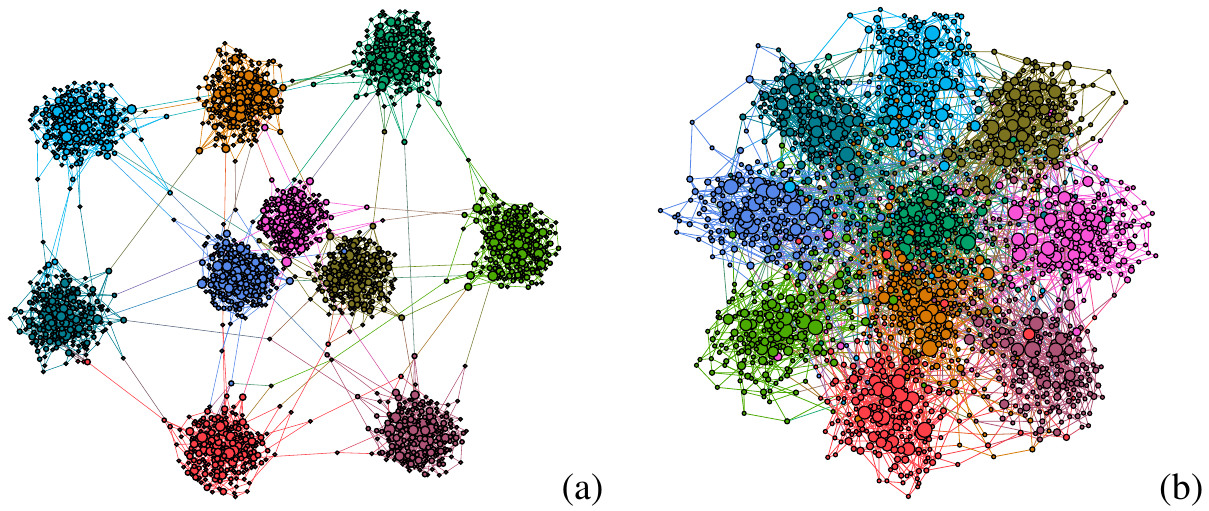}
	\caption{{Examples of modular structures}.
		Networks with $M=10$ modules of same size $S=200$ and number of intermodular
		connections (a) $\kout=10$  and  (b) 100  representing loosely
		and densely connected modular graphs, respectively. The network degree
		distribution is given by $P(k)\sim k^{-2.7}$ with $\klow=3$ and $\kupp=14$ for
		the lower and upper bounds cutoffs, respectively. Connected modular
		structures can clearly be observed. {Nodes in a same community 
			are plotted with the same color and their sizes are proportional to the vertex degree. The
			graph was generated using Gephi visualization tool (\href{https://gephi.org}{https://gephi.org}).}}
	\label{fig:modular_pout0}
\end{figure}

We generated modular networks based on the benchmark  model of Lancichinetti,
Fortunato, and Radicchi\cite{Lancichinetti2008}. Consider $g=1,\ldots,M$
modules where the size $S_g$ of each group is drawn according to a distribution
$Q(S_g)$. At a vertex level, the degrees are drawn from a distribution $P(k)$
with $k=\klow,\ldots,\kupp$ where $\klow$ and $\kupp$ are lower and upper
cutoffs of the degree distribution, respectively. The maximal number of \textit{intermodular}
edges connecting vertices of different groups is predefined as $\kout_g$ and, in
general, can depend on the module. By construction,  this  model produces highly
modular networks if the number of intermodular connections is much smaller than
the \textit{intramodular} one, which was confirmed by the calculation of the
modularity coefficient\cite{newman2010networks} and using the Louvain community
detection algorithm\cite{Blondel2008}. See section Methods
for network metrics and generation procedure. Figure~\ref{fig:modular_pout0} shows
examples of modular networks with different levels of intermodular connectivity.

Depending on the module size distribution $Q(S_g)$, we divided  the investigated
networks into two classes. In the monodisperse modular networks (MMNs), all
modules have exactly the same number of vertices and of intermodular
connections, \textit{i.e.} $S_g=S$ and $\kout_g=\kout$ for $g=1,\ldots, M$.
However, real modular networks are not monodisperse in the aforementioned sense
and, thus, we also considered a PL distribution of module size with $Q(S_g)\sim
S_g^{-\expS}$, which are consistent with observations in real
systems\cite{Lancichinetti2008}. The upper bound of the distribution is limited to
the system size ($S_g\le N$) while the lower one is chosen such that the average
module size $\av{S_g}$ has a predefined value. These networks are referred
hereafter as polydisperse modular networks (PMNs). We also chose the number of
intermodular connections proportional to the module size, $\kout_g\propto S_g$,
constraining $\kout_g\ge 2$, to guarantee connectivity\cite{Albert2002}. We used
the values of $\av{S_g}=10^3$  and $\av{\kout_g}=5$ in all presented results to
perform comparison between monodisperse and polydisperse cases. {
	Brain networks, which inspired our research, are not fully random like those presented 
        here, but our aim was to isolate the role played by the network dimensionality.}
    
\begin{figure}[hbt]
\centering
\includegraphics[width=0.6\linewidth]{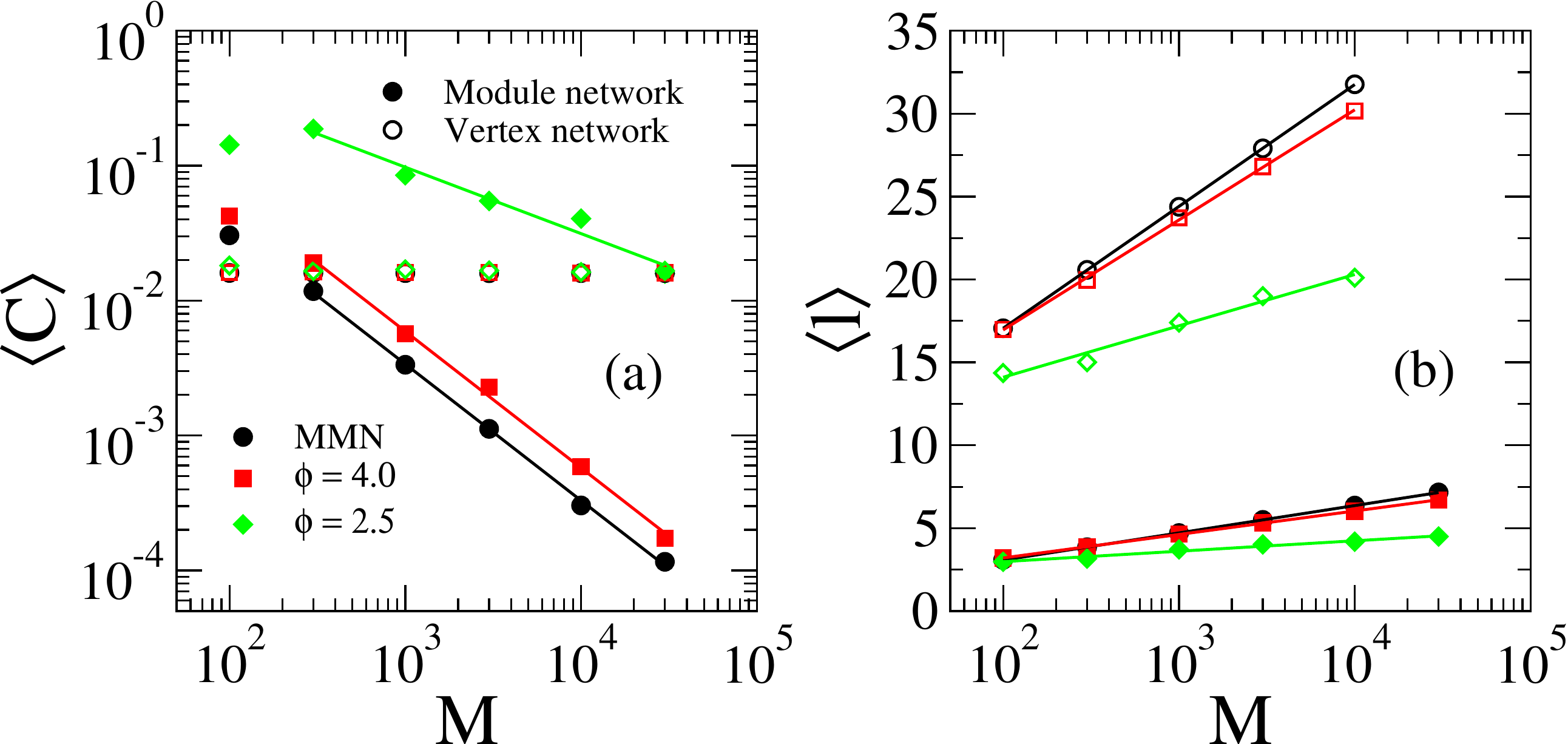}	
\caption{{Structural properties of modular networks}. (a) Clustering
	coefficient and (b) average shortest path as functions of the number of
	modules. Open symbols correspond to the vertex network, while the filled ones
	represent the network of modules, in which the modules are themselves treated
	as nodes connected by the intermodular edges. Lines denote (a) power-law  or
	(b) logarithmic regressions.  In monodisperse modular networks (MMN), all
	modules have the same number of vertices $S=\av{S_g}$. Networks with module
	size distribution $Q(S_g)\sim S_g^{-\expS}$ are obtained by fixing the minimal
	size $\Smin$  such that the chosen average size is obtained. The parameters are
	$\av{\kout_g}=5$, $\av{S_g}=10^3$, $\gamma=2.7$, $\klow=3$, and $\kupp=58$; see
	main text. { The averages were performed over 25 
		independent networks. }}
\label{fig:modular-statistics}
\end{figure} 

\begin{table}[hbt]
\begin{center}
	\begin{tabular}{ccc}
		\hline\hline
		& Vertex networks              & Module networks  \\ \hline
		MMN              & ~$\av{l} = 2.35 + 3.19\ln(M)$ & ~$ \av{l} = -0.22 + 0.71\ln(M)$   \\
		$\expS=4.0$   & ~$\av{l} = 3.67 + 2.88\ln(M)$ & ~$ \av{l} = 0.37 + 0.61\ln(M)$    \\
		$\expS=2.5$   & ~$\av{l} = 7.91 + 1.35\ln(M)$ & ~$ \av{l} = 1.73 + 0.272\ln(M)$    \\ 
		\hline \hline
	\end{tabular}
\end{center}
\caption{{Logarithmic regressions for the average shortest distance in modular
		networks}. We analyzed both the original vertex network and the one where
	modules are considered as vertices connected only by the intermodular edges. {Correlation
		coefficient of the regressions is $r^2>0.999$ for MMN and $\phi=4.0$, while $r^2=0.99$
		for $\phi=2.5$.}}
\label{tab:path}
\end{table}

We determined the average clustering coefficient and the average shortest mean
distance\cite{Albert2002} for both vertex and module networks. The latter means
that we treat modules as vertices, connected by intermodular edges forming a
network. Structural properties of these modular networks  are shown in
Fig.~\ref{fig:modular-statistics}. The clustering coefficient, averaged over the
whole network, saturates at a small finite value as the network size increases
(see Fig.~\ref{fig:modular-statistics}(a)). This is a natural consequence of the
modular organization of the network that forces vertices to be connected mostly
within the modules which are of finite size and thus the probability to form
triangles is not  negligible. The clustering coefficient of the network of
modules vanishes as $\av{C}\sim M^{-1}$ in the cases of $\expS=4.0$ and MMNs, while it
vanishes as $\av{C}\sim M^{-1/2}$ for $\expS=2.5$. Hierarchically organized
networks are clustered with coefficient independent of the
size\cite{Ravasz2003}. So, the analysis of Fig.~\ref{fig:modular-statistics}
shows the lack of hierarchy in the modular networks of our investigation.

The average shortest path is defined as the average minimal graph distance among
every pair of vertices\cite{newman2010networks}. For the presented modular
networks, this increases logarithmically with the size as shown in
Fig.~\ref{fig:modular-statistics}(b) and Table~\ref{tab:path}. So, the
investigated networks have infinite dimension besides the lack of hierarchy.

\subsection{Epidemic process on modular networks}

We ran time dependent simulations of SIS dynamics, in which
an infected (active) vertex $i$ spontaneously heals (inactivates) with rate
$\mu_i$ and infects each of its susceptible nearest neighbors with rate
$\lambda_i$. We used the statistically exact and optimized Gillespie algorithm
detailed elsewhere\cite{Cota2017} with different initial conditions: decay from
fully infected initial states and spreading simulations started from a single
infected vertex\cite{marro2005}; see Methods' section. Finite size effects were
investigated using networks of size $N\approx M\av{S_g}$ with $M=10^3$, $10^4$ and
$3\times 10^4$ modules, remembering that $\av{S_g}=10^3$ was adopted in all cases.

In the SIS dynamics, the transition point is governed by the long-term self
activation of hubs and their mutual reactivation through connected
paths\cite{chatterjee2009,Boguna2013,Ferreira2016a}, such that it presents a
null threshold in case of PL networks  $P(k)\sim k^{-\gamma}$ in the infinite
size limit with $\kupp\rightarrow\infty$.  In order to deal with a finite
threshold in the thermodynamic limit, we considered two types of disorders
{called hereafter \textit{topological} or \textit{intrinsic} disorder}. The
vertex degrees are distributed according to a truncated PL  with $\klow=3$,
$\kupp=58$ and $\gamma=2.7$ in the case of topological disorder. Infection and
healing rates $\lambda_i=\lambda$ and $\mu_i=1$ (fixing the time scale) are
uniform for all edges and vertices, respectively, and the disorder is due to
vertex degree variability. We considered $P(k)=\delta_{k,4}$ for the intrinsic
disorder case, such that each module forms a random regular network
(RRN)\cite{Ferreira2013}, in which every vertex has the same degree but the
connections are random. Since topological disorder is negligible in this RRNs,
the intrinsic disorder is introduced in the healing rates $\mu_i$ of each vertex
$i$ that take binary values $1-\epsilon$ or $1+\epsilon$ with equal chance,
while the infection rate is still uniform with $\lambda_i=\lambda$. Note that, 
when investigated on homogeneous degree networks such as RRNs,  the SIS is
equivalent to the contact process\cite{Harris74} used in previous studies of GPs
on networks\cite{Munoz2010,Juhasz2012,Odor2015,Odor2012}.

\subsubsection{Density decay analysis}

\begin{figure}[b!]
	\centering
	\includegraphics[width=0.72\linewidth]{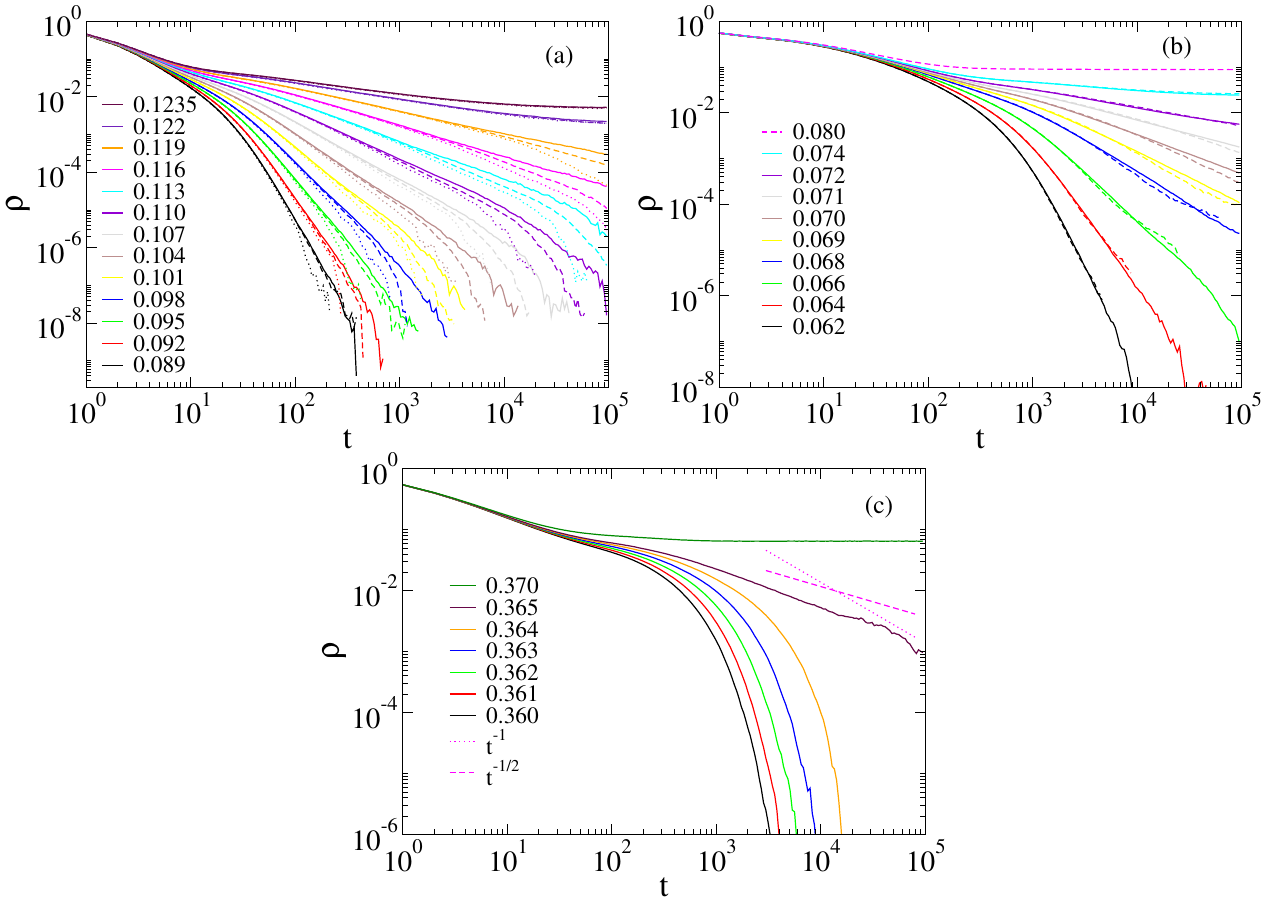}
	\caption{{Density decay on a single MMN}. (a) Decay analysis for SIS with
		topological disorder introduced by a degree 	distribution $P(k)\sim k^{-2.7}$
		and $\kupp=58$. The numbers of modules are $M = 10^3$ (dotted lines), $M =
		10^4$ (dashed lines), and $M = 3\times 10^4$ (solid lines). Legends indicate
		the values of $\lambda$. (b) Decay analysis for SIS with  intrinsic disorder
		($\epsilon=0.9$) on MMNs of sizes $M=10^3$ (dashed lines) and $M=10^4$ (solid
		lines) where the modules are themselves RRNs.  (c) SIS decay without intrinsic
		disorder ($ \epsilon=0$) on a single MMN of $M=10^3$ modules, each one consisting of
		a RRN.}
	\label{fig:homog-natlocal-M1k10k-sis}
\end{figure}

We show the density decays for a given realization of a  MMN  for three models
of disorder in Fig.~\ref{fig:homog-natlocal-M1k10k-sis}. Similar results were
found for the other analyzed network realizations (up to 20). For the
topological disorder, a finite size analysis increasing the number of modules is
presented in Fig.~\ref{fig:homog-natlocal-M1k10k-sis}(a). The curves reveal
non-universal PLs in the $0.089 \le \lambda < 0.12$ extended region, which do
not change within statistical error margins as the number of modules increases
from $M=10^3$ to $3\times 10^4$. Thus, contrary to the case of SIS on
non-modular PL networks\cite{Cota2016}, we see a GP behavior. It is important to
mention that the critical regimes hold for intermediate times since the networks
are still finite. Furthermore, the analysis provides numerical evidences that
the transition point is also size independent. The case of strong intrinsic
disorder given by $\epsilon=0.9$, shown in
Fig.~\ref{fig:homog-natlocal-M1k10k-sis}(b), also presents extended region of
critical behavior with non universal PLs preserved as the sizes are increased.
It is worth noting that the SIS dynamics on MMNs without intrinsic nor
topological disorder ($\mu_i=1$), shown in
Fig.~\ref{fig:homog-natlocal-M1k10k-sis}(c), does not show GPs and the critical
behavior is given by $\rho\sim t^{-1/2}$, instead of a regular mean-field
decay\cite{Henkel2008} $\rho\sim t^{-1}$. This was also found in generalized
small-world networks for which the GP shrank to a very narrow
region\cite{Juhasz2012}. We also investigated weaker intrinsic disorder using
$\epsilon=0.5$ and observed GPs in several networks realizations, but in others
they were weak or absent. However, when we performed disorder realization
averaging, GPs became evident for both values of $\epsilon$ (see Supplementary
Information for figures).

\begin{figure}[b!]
	\centering
	\includegraphics[width=0.72\linewidth]{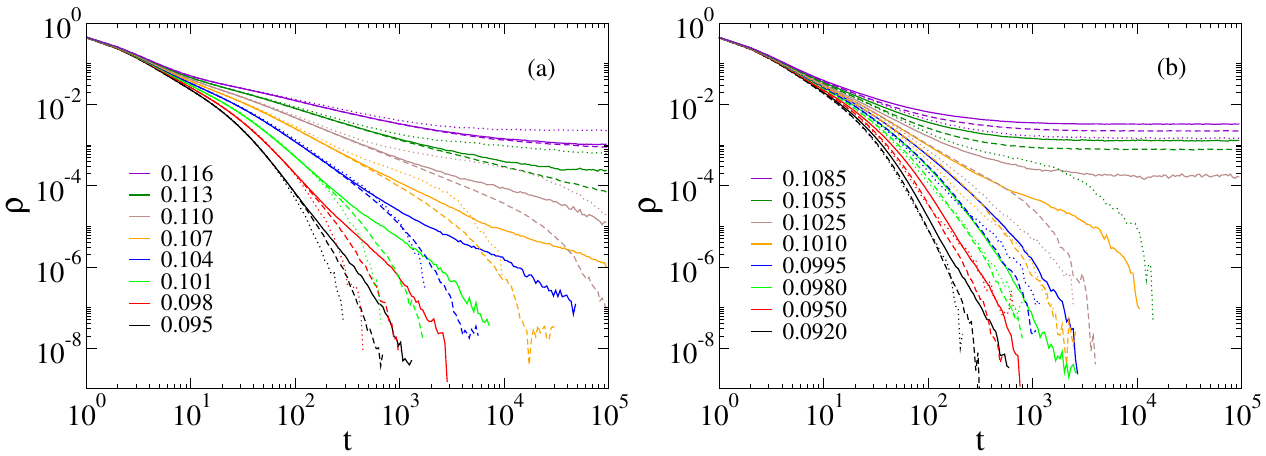}
	\caption{{Density  decay for SIS with topological disorder on a single
			PMN.}  The module size distributions with exponents 	(a) $\expS = 4.0$ and (b)
		$\expS = 2.5$ are shown. The finite size analysis is done using  $M = 10^3$
		(dotted lines), $10^4$ (dashed lines), and $3\times 10^4$ (solid lines). Other
		network parameters are given in text and the values of $\lambda$ indicated in
		the legends. }
	\label{fig:400-rigid58-M1k10k-sis}
\end{figure}

{We investigated the effects of module size variability considering PMNs 
        with topological disorder only. These exhibit the same truncated PL for degree
	distribution and average sizes of modules as those of MMNs to permit a comparison.} In
Figure~\ref{fig:400-rigid58-M1k10k-sis}(a), we show extended regions of
$\lambda$ with PL tails in the density time decays for $\expS=4.0$, which
corresponds to a heterogeneous, but finite variance distribution. These results
look qualitatively similar to those of the MMN case. Finite size effects are
stronger, but a GP occurs in the interval $0.095<\lambda<0.115$, which is
narrower than in the monodisperse case. Noticeably, GPs are not observed for the
scale-free case with $\expS=2.5$ shown in
Fig.~\ref{fig:400-rigid58-M1k10k-sis}(b), in which modules of essentially every
size appear. A finite variance of $Q(S_g)$ reduces the RR effects in comparison
with MMNs, since some large modules have many intermodular connections $\kout\gg
\av{\kout}$ reducing their independence. For an infinite variance the situation
becomes drastic. A single module can contain a considerable large fraction of
the whole network and alone rules the critical dynamics of the system becoming
equivalent to the non-modular case\cite{Cota2016,odor13a,odor13b}.
{Once we have established  under which conditions of module size
	variability the GPs are robust, in the rest of the paper we consider only the
	case of MMNs, stressing that the central conclusions are the same as in the case
	of a finite variance in the module size distribution.}

\subsubsection{Spreading analysis}

\begin{figure}[hbt]
	\centering
	\includegraphics[width=0.72\linewidth]{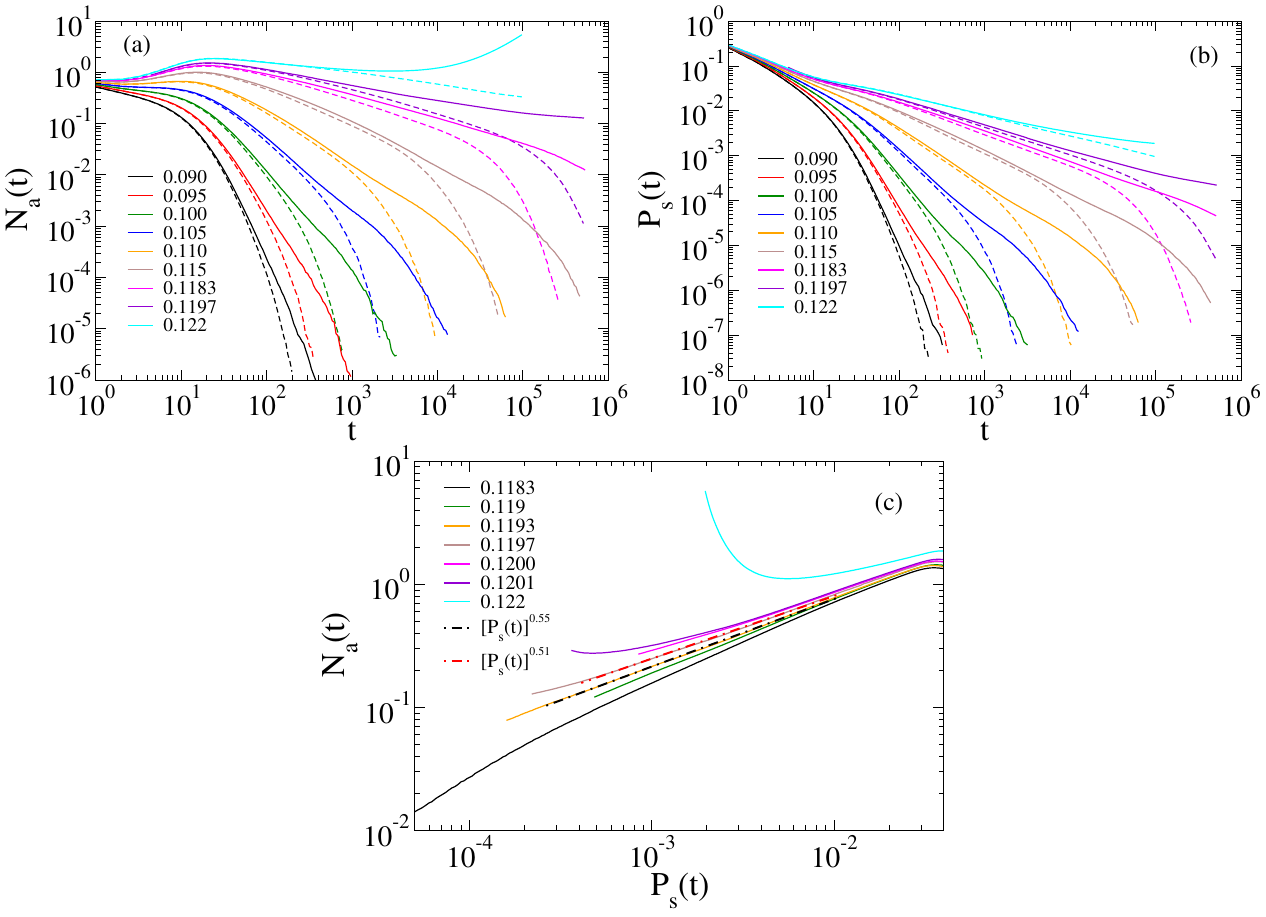}
	\caption{{Spreading analysis for SIS on a single MMN}. Only topological
		disorder is introduced as a truncated PL $P(k)\sim k^{-2.7}$ with $\kupp=58$.
		Finite size analyses of the (a) number of active nodes and (b) survival
		probability are done with  $M=10^3$ (dashed lines) and $10^4$ (solid lines)
		modules. (c) Determination of the transition point in a double logarithmic plot
		of $N_\text{a}(t)$ vs $P_\text{s}(t)$ for $M=10^4$ modules. Infection rates $\lambda$ are indicated in the legends.} 
	\label{fig:spread}
\end{figure}

Fig.~\ref{fig:spread}(a) shows the number of active vertices as a function of
time in spreading simulations on a MMN. For regular dynamical criticality, this
quantity is expected to evolve as $N_\text{a}(t) \propto t^{\eta}$. One can see
non-universal PL tails in a range similar to the one found in the density
decays{, including a similar exponential cutoff for long times due to
	the finite size of the networks.}. The survival probability curves,
$P_\text{s}(t)$, exhibit a very similar behavior [Fig.~\ref{fig:spread}(b)] with
the same exponents as those of the $\rho(t)$ decays at a given $\lambda$ within
the critical region, expressing that the rapidity reversal
symmetry\cite{Henkel2008,odor04} is unbroken by the quenched disorder. This
symmetry implies, for example, that the asymptotic probability
($t\rightarrow\infty$) to find one infected vertex at a randomly chosen location
is weakly  dependent on the initial condition or, more precisely,
$\rho(t)\propto P_\text{s}(t)$.

Due to the extended interval with PLs and the corrections, it is hard to
estimate the transition point location and the time decay functional form
accurately. Simple PL fitting results in $P_s(t) \propto t^{-\delta}$ with
$\delta=0.42(1)$ at $\lambda_\text{c}\simeq 0.12$. Assuming a scaling in the form
	$P_s(t) \sim \ln(t/t_0)^{-\tilde\delta}$, as in case of the absorbing state
	phase transition with strong disorder in lower dimensions\cite{Moreira1996}, we
	could obtain $\tilde\delta\approx 5$. Neither of these is in agreement with the
regular mean-field behavior obtained for absorbing state phase transitions
with quenched disorder in high dimensions\cite{Vojta2014}. We applied an
alternative method\cite{Juhasz2012}, which assumes that the leading correction
to the scaling comes from the same scale $t_0$ in the critical behaviors of
$P_\text{s}(t)$ and $N_\text{a}(t)$. Plotting $\ln[N_\text{a}(t)]$ against
$\ln[P_\text{s}(t)]$, transition point curves must fit on a straight line. As
Fig.~\ref{fig:spread}(c) shows, this allows an estimate for the transition point
$\lambda_\text{c}=0.1195(2)$, in which the slope is $0.53(2)$.

\begin{figure}[ht]
	\centering \includegraphics[width=0.45\linewidth]{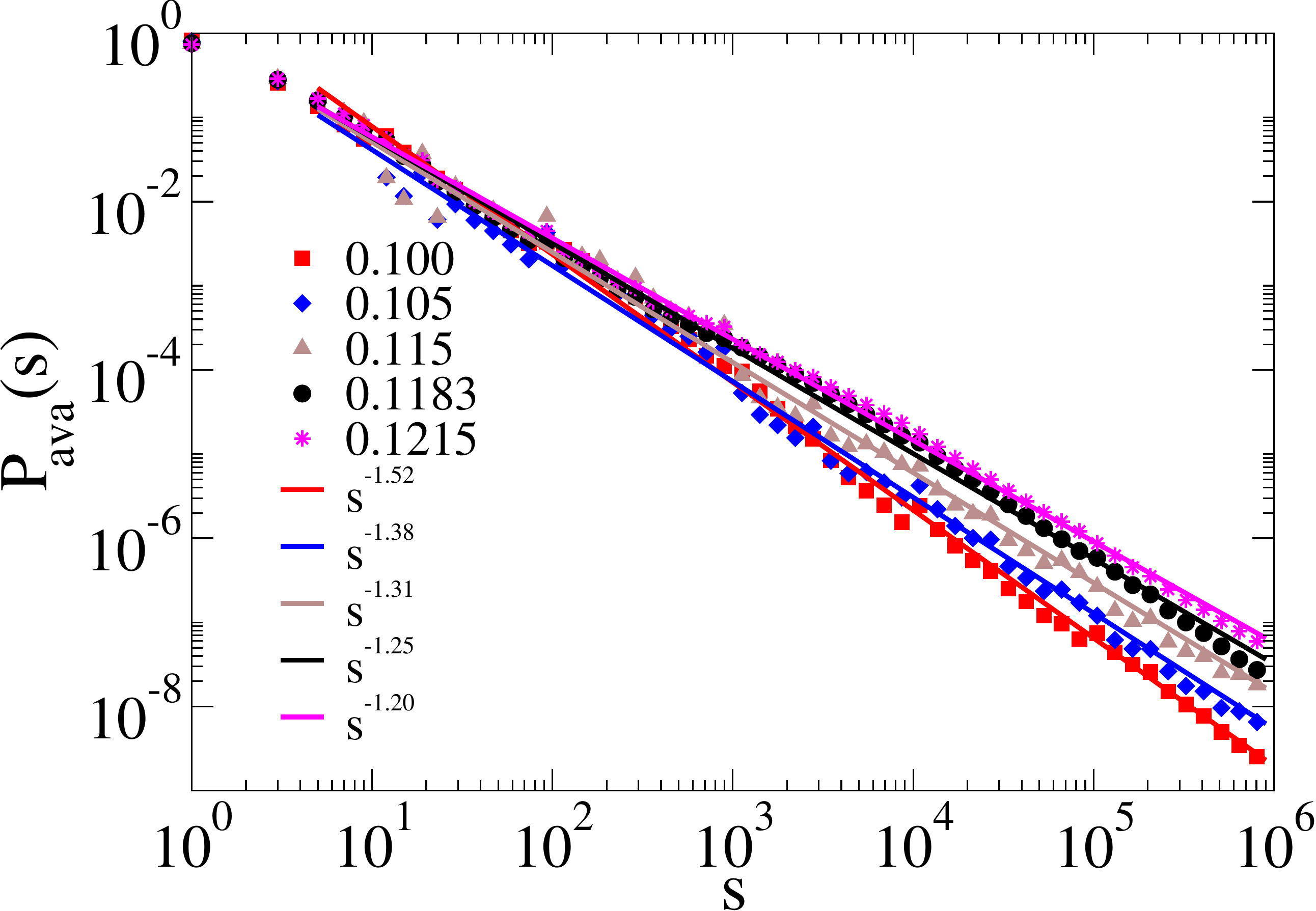}
	\caption{{Avalanche size distributions of the SIS spreading on a single MMN.}
		Only topological disorder is considered and  the number of modules is $M=10^3$.
		Different values of $\lambda$ are indicated by the legends. Simple PL tail fits
		are also shown.} \label{spha}
\end{figure}

We also determined the avalanche size distributions $P_\text{ava}(s)$ in
spreading simulations. The size of an avalanche is defined as the total number
of  sites $s$ activated during a spreading experiment. The results for $M=10^3$
can be seen in Fig.~\ref{spha}. Power-law behavior occurs for the $10 < s <
10^6$ region with a variation of the exponent as a function of $\lambda$. A PL
fitting for the $0.1 \le\lambda\le 0.1215$ region results in
$P_\text{ava}(s)\propto s^{-\tau}$ with $1.20 \le \tau \le 1.52$, which
{encloses mean-field exponent} of the directed percolation class
($\tau=3/2$)\cite{Munoz99}. Curiously, this mean-value is consistent with
reports for activity avalanches observed in the brain\cite{Beggs2003,Beggs2012}.
However, it is important to remember that other mechanisms can explain the
exponent\cite{Chialvo2010} $\tau=3/2$. {On the other hand, since the
	spreading and decay exponents depend on $\lambda$, the same should happen for
	$\tau$, as the consequence of the scaling relation $\tau = (1+\eta+2\delta) / (1+\eta+\delta)$ 
	for absorbing state phase transitions~\cite{Munoz99}. Indeed, the  3/2
	exponent  also appears in the avalanche mean-field exponents of many models and
	several universality classes (DP, Dynamical Percolation, Random Field Ising
	Model, etc.)~\cite{odor04}. However, in our case, the universality class seems
	to be different, since the 3/2 exponent is found within the GP, while at the
	critical point $\lambda_c\approx 0.12$ the measured $\tau$ is smaller. 
	}

\section{Discussion}

{
	We have analyzed the activity spreading of the SIS model in loosely connected,
non-hierarchical modular networks and observed extended regions of
non-universal scaling behavior for intrinsic
[Figs.~\ref{fig:homog-natlocal-M1k10k-sis}(a),
\ref{fig:400-rigid58-M1k10k-sis}(a) and \ref{fig:spread}] and topological
[Fig.~\ref{fig:homog-natlocal-M1k10k-sis}(b)] disorders, using density decay
and spreading analysis. The interval of the control parameter with dynamical
critical region, which was robust under finite size analysis, showed size similar
to previous studies on hierarchical
networks\cite{Moretti2013,Li2017,Odor2015}. It depends on the disorder type and
shrinks only in case of a module size distribution of diverging variance
[Fig.~\ref{fig:400-rigid58-M1k10k-sis}(b)]. The time window where we observed
scaling undergoes an exponential cutoff. The reason is that we are dealing with
infinite dimensional networks, for which increasing the number of vertices by
one order of magnitude increases the diameter only by a few unities as shown in
Fig.~\ref{fig:modular-statistics}(b). However, the range of the scaling regime
is improved with the size, as can clearly be read off from
Fig.~\ref{fig:homog-natlocal-M1k10k-sis}(a). This is clarified further via the
local slope analysis in the Supplementary Information. However, the time window
size of power laws increases modestly for non MMNs. Note that 
logarithmic corrections in GPs are common and this can really be observed by our local
slope analysis shown in the Supplementary Information. Finally, the presented
scaling regimes correspond to single network realizations, while the power-law
regime is increased  if  averages are performed  over many independent networks
as exemplified in the Supplementary Information.
}

At a first glance, the results presented up to this point are in odds with the
conjecture that infinite dimensional networks cannot sustain real
GPs\cite{Munoz2010}. Strictly speaking, one may argue that the finite module
sizes in the monodisperse case imply that RR lifespans can be huge, but bounded
and thus the observed PLs correspond to very strong Griffths effects, differing
from real GPs in the sense that they disappear in the thermodynamic limit.
However, we also found size-independent GPs in the polydisperse case with
$\expS=4.0$, where this size restriction does not apply. To understand this, we
express the typical $S_c$ for which at least one module of size $S>S_c$ is
present as
\begin{equation}
M\int_{S_c}^{\infty}Q(S)dS\sim 1 \Rightarrow S_c\sim M^{1/(\expS-1)},
\end{equation}
implying that the sizes of the largest modules diverge as their number is
increased. The same result can be deduced more rigorously using extreme-value
theory\cite{Boguna2004a}.

\begin{figure}[hbt]
	\centering
	\includegraphics[width=0.45\linewidth]{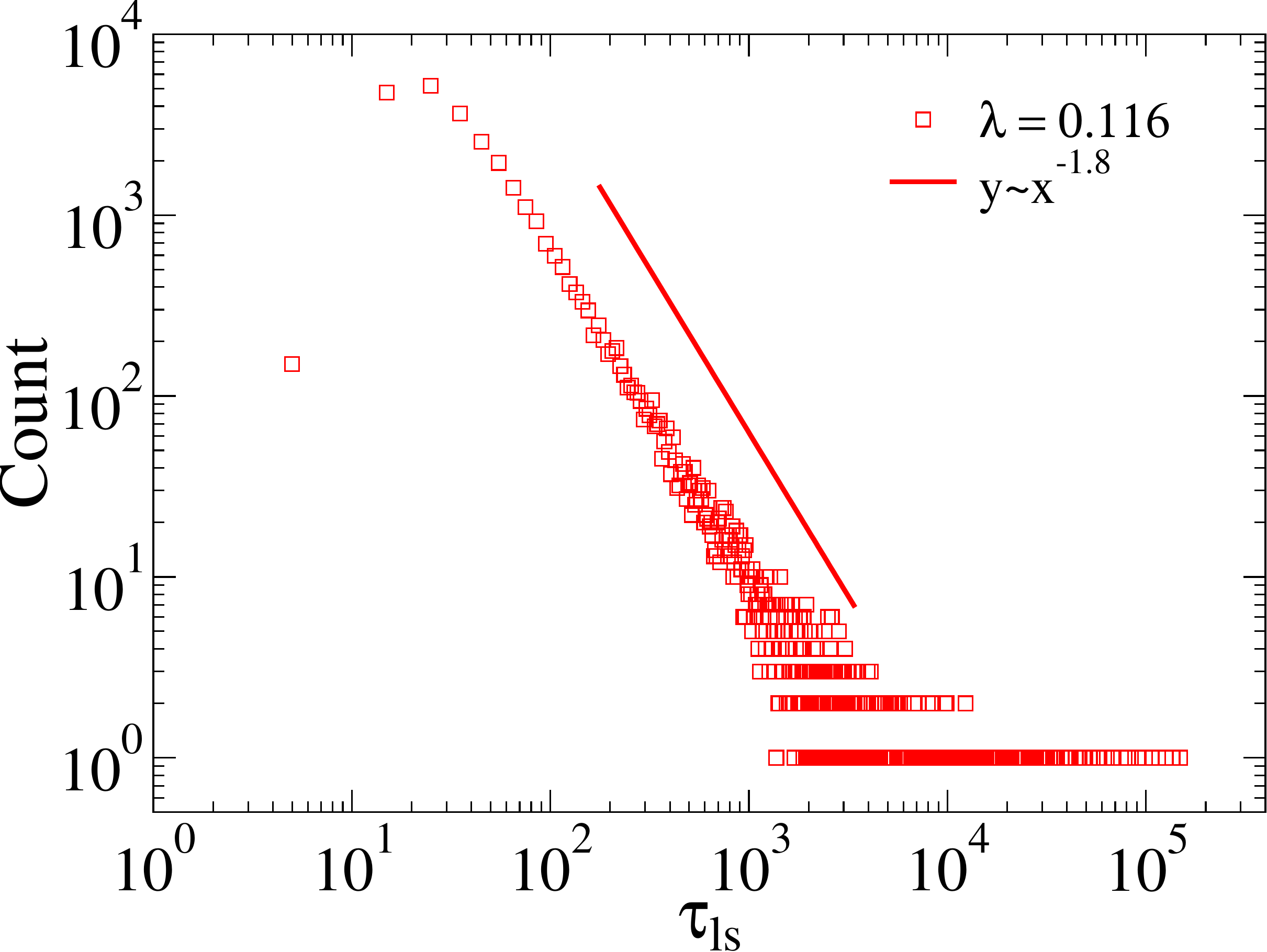}
	\caption{{{Distribution of activity lifespan within individual
				modules}. Distribution of the activity lifespan $\tau_\text{ls}$ for SIS
			dynamics computed in individual modules of a single MMN with $M=3\times 10^4$ and 
			$\lambda=0.116$ inside the GP. The distribution is computed by binning  
			time windows of size $\Delta \tau_\text{ls} = 10$. }} \label{fig:decay_modules_HMM_30k}
\end{figure}

The underlying mechanism behind GPs is {the emergence of RRs related to
} the nature of the activity spreading of the SIS model and to the loose
intermodular connectivity {of the network}. Let us discuss
{this in the case of} topological disorder, but similar reasoning
{can be applied} for the intrinsic one too. {Activity of the
	SIS dynamics is concentrated in localized {(sub-extensive)} domains,
	which can be hubs\cite{Castellano10,Mata2015,Lee2013} or densely connected
	groups of vertices in the innermost core of the
	network\cite{Kitsak2010,Castellano2012}. However, bridges among modules
	{in the investigated model} are randomly built, implying that the
	probability of being connected through highly active regions of different
	modules is very small. This is associated with the variability of module
	properties, caused by the randomness. This produces broad activity lifespan
	distributions in individual modules inside the GP,
	$\lambda_0<\lambda<\lambda_c$, as shown in Fig.~\ref{fig:decay_modules_HMM_30k}
	for a MMN. In this case, the distribution follows $P(\tau_\text{ls})\sim
	\tau_\text{ls}^{-a}$, with an exponent $a\approx 1.8$. To obtain this law we
	computed the density of infected vertices as function of time in each
	module. The results could be fitted with an exponential decay $\rho\sim
	\exp(-t/\tau_\text{ls})$ in the range $t>20$ and we could extract the characteristic time
	$\tau_\text{ls}$ of each module individually. Active sites in a large fraction
	of these modules are therefore short lived and, from the point of view of the
	spreading process, behave as if they were removed. The remaining network that
	sustains the long-term activity can be approximated by isolated or weakly
	coupled patches, providing an effective zero dimensional substrate for the
	activity spreading.}
However, it must be stressed that there exist intermodular interactions, that
change the decay profiles in comparison with the isolated module case (see
Supplementary Information for plots).

To summarize, our analysis reveals the existence of stable GPs on small-world,
thus infinite dimensional substrates, conditioned to be sparsely connected in a
modular structure, as an alternative for the origin of criticality on modular
systems. The hierarchical modular networks, where GPs were previously
observed\cite{Moretti2013,Odor2015} are loosely connected. Hierarchy plays an
important role, by increasing the distances between the modules, thus enhancing
GPs, but it is not a necessary condition. The brain criticality hypothesis via
GPs raised by Moretti and Mu\~noz\cite{Moretti2013} is strengthened by our
results {since building  connectome
	networks~\cite{Hilgetag2016,sporns2010networks,Gastner2016} is far from of being
	trivial and the hypothesis of hierarchy in the modular organization of the brain
	is not fully accepted. Especially the very restrictive condition of finite
	dimension is fragile due to the presence of long-range connections. The
	model we used, conceived to be simple, allowed us to address this specific
	issue which would not easily be accessed in a real brain network.} We expect that
our results, which were not conceived for a specific system, will be important
for the investigation of criticality in other modular systems {beyond brain
	networks}. In the future, one should investigate real networks with the
aforementioned properties and build models with more realistic features, such as
correlation patterns\cite{Vazquez2002}, exerting significant influence on the
results.

\section{Methods}

\subsection{Generation of synthetic modular networks}

The network is generated as follows:
\begin{enumerate}[label=(\roman*)]
	\item The number of stubs of each vertex is drawn according to the degree distribution $P(k)$. 
	\item Two stubs  are randomly chosen.  If they belong to
	the same group, a new edge is formed.  If not, an edge is formed only if the
	maximal number of intermodular connections in both groups is not exceeded.
	\item Multiple or self-connections are forbidden.
	\item The process is iterated until all stubs are connected or it becomes
	impossible to form new edges without multiple or self-connections.
	\item The unconnected stubs are removed. We study only  
	the giant component which, in the present studies, 
        contains almost all vertices of the network.
\end{enumerate}
{The number of removed stubs is  a tiny fraction (less than 0.02\% of
	the stubs) and does not play any relevant role on the network properties shown
	in Fig.~\ref{fig:modular-statistics} or Table~\ref{tab:path}.
}
\subsection{Network metrics}

The modularity coefficient is defined by~\cite{newman2010networks}
\begin{equation}
Q_\text{mod}=\frac{1}{N\av{k}}\sum\limits_{ij}\left(A_{ij}-
\frac{k_ik_j}{N\av{k}}\right)\delta(g_i,g_j),
\end{equation}
where $A_{ij}$ is the adjacency matrix, defined as $A_{ij}=1$, if vertices $i$
and $j$ are connected and $A_{ij}=0$ otherwise; $\delta(i,j)$ is the Kronecker
delta function and $g_i$ corresponds to the community that vertex $i$ belongs
to. For $\kout\ll M$, we find $Q_\text{mod}\approx 1$, confirming the expected modular
structure of the investigated synthetic networks.  

The Watts-Strogatz clustering coefficient of a vertex $i$ is defined
as~\cite{newman2010networks}
\begin{equation}
C_i=\frac{e_i}{k_i(k_i-1)/2},
\end{equation}
where $e_i$ is the number of edges interconnecting the $k_i$ nearest
neighbors of node $i$. The average clustering coefficient is a simple average over all vertices
of the network.

\subsection{Computer implementation of the SIS model}

\newcommand{\Ninf}{N_\mathrm{inf}}
\newcommand{\Ne}{N_\mathrm{e}}

Statistically exact simulations of the SIS dynamics in a network with 
infection and healing rates $\lambda$ and $\mu$ can efficiently be 
simulated using pseudo-process method described elsewhere\cite{Cota2017}. A list with all infected 
vertices, their number $\Ninf$ and the number of edges
$\Ne$ emanating from them are recorded and constantly updated. 
In each time step, we proceed as follows. 
\begin{enumerate}[label=(\roman*)]
\item With probability
\begin{equation}
p = \frac{\mu \Ninf}{\mu \Ninf+\lambda \Ne}
\end{equation}
an infected vertex is selected with equal chance and healed.
\item With complementary probability $1 - p$, an infected vertex is selected
with probability proportional to its degree. A neighbor of the selected vertex
is chosen with equal chance and, if susceptible, it is infected. Otherwise
no change of state happens and the simulation runs to the next time step.
\item  The time is incremented by
\begin{equation}
\tau = -\frac{\ln(u)}{\mu \Ninf + \lambda \Ne}
\end{equation}
where $u$ is a pseudo random number, uniformly distributed in the interval $(0,1)$.
\end{enumerate}

For each network realization averages were computed over 100 to 500
independent dynamic runs in the decay simulations, started with all vertices
infected. For spreading analyses, which begins with a single infected vertex, the
process is started 10 to 100 times at each vertex of the network.

\subsection{Data Availability}

The datasets generated during and/or analysed during the current study are available from the corresponding author on reasonable request.

\bibstyle{naturemag-doi}
{\small \bibliography{gpSISmodular}}

\section{Acknowledgements}

This work was partially supported by the Brazilian agencies CAPES, CNPq and
FAPEMIG and the Hungarian research fund OTKA (K109577). We thank Robert Juh\'asz
for fruitful comments and discussions. G.\'O. thanks the Physics Department at
UFV, where part of this work was done, for its hospitality. S.C.F. thanks
the support from the program  \textit{Ci\^encia sem Fronteiras} - CAPES under project No. 
88881.030375/2013-01.

\noindent
This is a post-peer-review,
pre-copyedit version of an article published in Scientific Reports. 
The final authenticated version is available online at:
\href{http://dx.doi.org/10.1038/s41598-018-27506-x}{http://dx.doi.org/10.1038/s41598-018-27506-x}.

\section{Author contributions statement}

G.\'O. and S.C.F. conceived the research and wrote the paper. W.C. performed
dynamical simulations and structural analyses. S.C.F. generated the networks.
G.\'O. and S.C.F.  performed spreading simulations. All authors reviewed the
manuscript and analyzed the data.

\section{Additional information}

\noindent\textbf{Competing financial interests} The authors declare no competing interests.

\medskip

\noindent\textbf{Supplementary information} accompanies this paper at \href{http://dx.doi.org/10.1038/s41598-018-27506-x}{http://dx.doi.org/10.1038/s41598-018-27506-x}.

\end{document}